\documentclass[]{spie}  

\usepackage{amsmath,amsfonts,amssymb}
\usepackage{graphicx}
\usepackage[colorlinks=true, allcolors=blue]{hyperref}

\title{Convolutional variational autoencoders for secure lossy image compression in remote sensing}

\author[a]{Alessandro Giuliano }
\author[a]{S. Andrew Gadsden}
\author[a,b]{Waleed Hilal}
\author[a,b]{John Yawney}

\affil[a]{Intelligent and Cognitive Engineering (ICE) Lab, McMaster, 1280 Main Street, Hamilton ON, Canada}
\affil[b]{Adastra Corporation, 200 Bay Street, Toronto ON, Canada}

\authorinfo{Further author information: (Send correspondence to Alessandro Giuliano)\\Alessandro Giuliano: E-mail: giuliana@mcmaster.ca \\ S. Andrew Gadsden : E-mail: gadsdesa@mcmaster.ca \\ Waleed Hilal : Email : hilalw@mcmaster.ca \\ John Yawney: Email: john.yawney@adastragrp.com}

\begin{document} 
\maketitle

\begin{abstract}

The volume of remote sensing data is experiencing rapid growth, primarily due to the plethora of space and air platforms equipped with an array of sensors. Due to limited hardware and battery constraints the data is transmitted back to Earth for processing. The large amounts of data along with security concerns call for new compression and encryption techniques capable of preserving reconstruction quality while minimizing the transmission cost of this data back to Earth. 
This study investigates image compression based on convolutional variational autoencoders (CVAE), which are capable of substantially reducing the volume of transmitted data while guaranteeing secure lossy image reconstruction. CVAEs have been demonstrated to outperform conventional compression methods such as JPEG2000 by a substantial margin on compression benchmark datasets. The proposed model draws on the strength of the CVAE's capability to abstract data into highly insightful latent spaces, and combining it with the utilization of an entropy bottleneck is capable of finding an optimal balance between compressibility and reconstruction quality. The balance is reached by optimizing over a composite loss function that represents the rate-distortion curve.

\end{abstract}

\keywords{Compression, Encoding, Encryption, Neural Compression, Variational Autoencoders}

\section{INTRODUCTION}
\label{sec:intro}  

The recent push in the private aerospace sector has provided more cost-effective opportunities for both companies and governments alike to launch new missions. The number of satellite launches per year has drastically soared as a result, increasing the number of satellites orbiting Earth from approximately 1,400 in 2015 to nearly 5,500 by the spring of 2022. This upward trajectory is projected to intensify, with numerous experts forecasting the launch of approximately 58,000 additional satellites by the close of the decade. Part of these satellite launches include imaging satellites, and recovering multispectral, radar and LIDAR observational data. Some of these missions include the Sentinel mission from the European Space Agency (Sentinel 1 A and B, Sentinel 2 A and B, Sentinel 3A and Sentinel 6A) \cite{potin_status_nodate}, as well as NASA's Landsat missions (Landsat 7-8-9)\cite{crawford_50-year_2023} and from the Canadian Space Agency (CSA) with the RADARSAT 2 mission \cite{caves_radarsat-2_nodate} to name a few\cite{mccafferty-leroux_improved_2023}.

Furthermore, new technological advances now facilitate commercial-off-the-shelf (COTS) cameras to serve as payloads for CubeSats destined for Earth observation purposes, widening the application scope of small-scale satellite missions. The satellite imaging domain is currently experiencing a transformative phase, characterized by technological innovations that not only bolster the capabilities of remote sensing but also enhance the efficiency and cost-effectiveness of global monitoring endeavors. Nevertheless, endeavors to downsize payload instruments to comply with the size, weight, and power (SWaP) limitations of small spacecraft persist. Limitations in these new class of satellite also include computational and hardware constrains. In the last five years, numerous CubeSat missions centered on Earth observation have been initiated to showcase optical imaging payloads that surmount these obstacles. Some of these small satellite missions include SeaHawk-1 and SeaHawk-2 from the University of North Carolina-Wilmington, Spectral Ocean Color (SPOC) by the University of Georgia and the BeaverCube-2 (BC-2) from the Massachusetts Institute of Technology (MIT) \cite{tomio_commercially_nodate}. Also the ESA's PhiSat-1 mission, which aims to demonstrate the feasibility of onboard processing by including a power-constrained machine learning accelerator and software for custom applications.  

The increased data availability is opening the door to novel applications of satellite imagery data including flood prediction \cite{twele_sentinel-1-based_2016}, land cover classification \cite{he_land_2018},  and environmental change detection \cite{chang_multisensor_2018}.

However, the transmission of large sensor images from CubeSats to ground receivers is limited by power and bandwidth constraints. Within this framework, on-board compression assumes a pivotal role in conserving transmission channel bandwidth and diminishing data-transmission duration, addressing memory and complexity constraints. In this context, compression methodologies can be classified into three principal categories: lossless, near-lossless, and lossy compression. Lossless compression is a reversible method that condenses data without sacrificing information, perfectly reconstructing the original data. The entropy metric, which gauges the information encapsulated within a source, establishes a theoretical threshold for lossless compression, such as the minimum attainable bitrate. Lossy compression, on the other hand, achieves higher compression ratios by discarding some information that may not be crucial for the application at hand while near-lossless represents the highest accuracy lossless algorithm approaching but not reaching lossless compression. 

To overcome these limitations, onboard processing has emerged as a viable solution to reduce the amount of data that needs to be transmitted \cite{tomio_commercially_nodate}. Variational Autoencoders (VAEs) have shown great potential in the field of image compression. VAE's can efficiently encode and decode images while preserving important features \cite{wu_review_2024}. By utilizing variational autoencoders for satellite image compression, it is possible to reduce large sensor images captured by CubeSats into smaller data products. These smaller data products, such as flood masks or compressed images, can be transmitted more efficiently within the bandwidth and power constraints of CubeSats. In this article we will cover lossy image compression using a lightweight VAE, and explain the rate-distortion curve to showcase the tradeoff between compressibility and distortion.

\section{Background on Satellite Image Compression}

Optical satellites equipped with payloads like panchromatic, multispectral, hyperspectral, Fourier transform spectroscopy (FTS), and light detection and ranging (LIDAR) capture data by measuring reflected light across different wavelengths. The advancement of sensor technologies has led to optical satellites with increased spectral bands, spatial resolution, swath, and radiometric precision to meet user and decision-maker demands. However, this progress has resulted in a significant increase in data volumes, especially from hyperspectral sensors that generate large image cubes. To effectively manage this vast amount of data, satellite image compression techniques are essential to reduce data size for efficient transmission, storage, and processing. In satellite applications where preserving data integrity is critical, near-lossless compression is of particular interest as it strikes a balance between data compression and maintaining scientific information necessary for remote sensing applications \cite{qian_optical_2013}. 

\subsection{Conventional Compression Methods for Satellite Hyperspectral Images}
\label{sec:title}

Conventional satellite compression methods focus on near-lossless compression to achieve high fidelity in relaying these images back to Earth. These methods generally use handcrafted transforms, quantization, and entropy coding techniques, such as the JPEG 2000 standard. In the specific context of satellite imaging, three key characteristics need to be fulfilled by image compression algorithms. First, quasi-lossless compression is necessary to maintain the accuracy and detail of satellite images for interpretation tasks on the ground. Second, the algorithms must be computationally efficient to accommodate the processing constraints onboard the satellite. Third, satellite images often contain small objects and high entropy, meaning that compression techniques need to preserve these high-frequency details. Vector quantization has been widely used for satellite image compression because of its ability to preserve fine details and spatial information. Recent advancements in satellite image compression have seen a shift towards the development of more advanced techniques that not only focus on near-lossless compression but also address the challenges specific to satellite imaging. One such challenge is the need for maintain the accuracy and detail of satellite images for interpretation tasks on the ground. This entails developing compression algorithms that can effectively preserve the high-frequency details present in satellite images, including small objects and regions with high entropy

JPEG (Joint Photographic Experts Group) is a widely used image compression standard that plays a significant role in satellite image compression. The JPEG compression method, based on near-lossless compression techniques, is particularly effective for reducing the size of images while maintaining visual quality. It achieves compression by analyzing and encoding image data in ways that allow for significant reductions in file size without perceptible loss in image quality, it can both render lossless and lossy compression.

JPEG2000 compression involves several key steps, including color space transformation, discrete cosine transform (DCT), quantization, and entropy encoding using embedded block coding with an optimized interception (EBCOT) \cite{yu_analysis_nodate}. Color space preprocessing converts the original RGB color space into the YCbCr color space, which separates brightness (luminance) information from color (chrominance) information. The DCT process then converts image blocks from the spatial domain to the frequency domain, allowing for more efficient representation of image data. Quantization reduces the precision of the DCT coefficients and EBCOT performs binary arithmetic to further compress the data. It is worth noting that entropy encoding is also often performed using Huffman coding, generating the compressed JPEG image file with optimal bit coding but without optimizing the rate-distortion criterion \cite{yu_analysis_nodate}.

Other types of near-lossless image compression techniques used on board of satellites include the multi-stage vector quantization (SAMVQ) and cluster vector quantization (HSOCVQ) algorithms \cite{qian_near_2006}. These methods are designed to control compression errors to a level comparable to the intrinsic noise in the original datasets. By setting compression fidelity thresholds below the intrinsic noise level, these techniques aim to minimize the impact of compression errors on remote sensing applications compared to the intrinsic noise. SAMVQ operates by organizing continuous 2D focal plane frames into regional datacubes. It splits a regional datacube into subsets for parallel processing, classifying them based on the similarity of spectra within the datacube. This approach allows for the independent compression of subsets, improving processing speed and memory utilization. HSOCVQ on the other hand classifies a regional datacube into clusters based on the similarity of spectra rather than dividing it into vignettes. This method is advantageous because it groups similar spectra into clusters, enhancing the compression process by associating spectra with the same clusters of specific targets in the scene \cite{qian_optical_2013}. 

   \begin{figure} [b]
   \begin{center}
   \begin{tabular}{c} 
   \includegraphics[height=7cm]{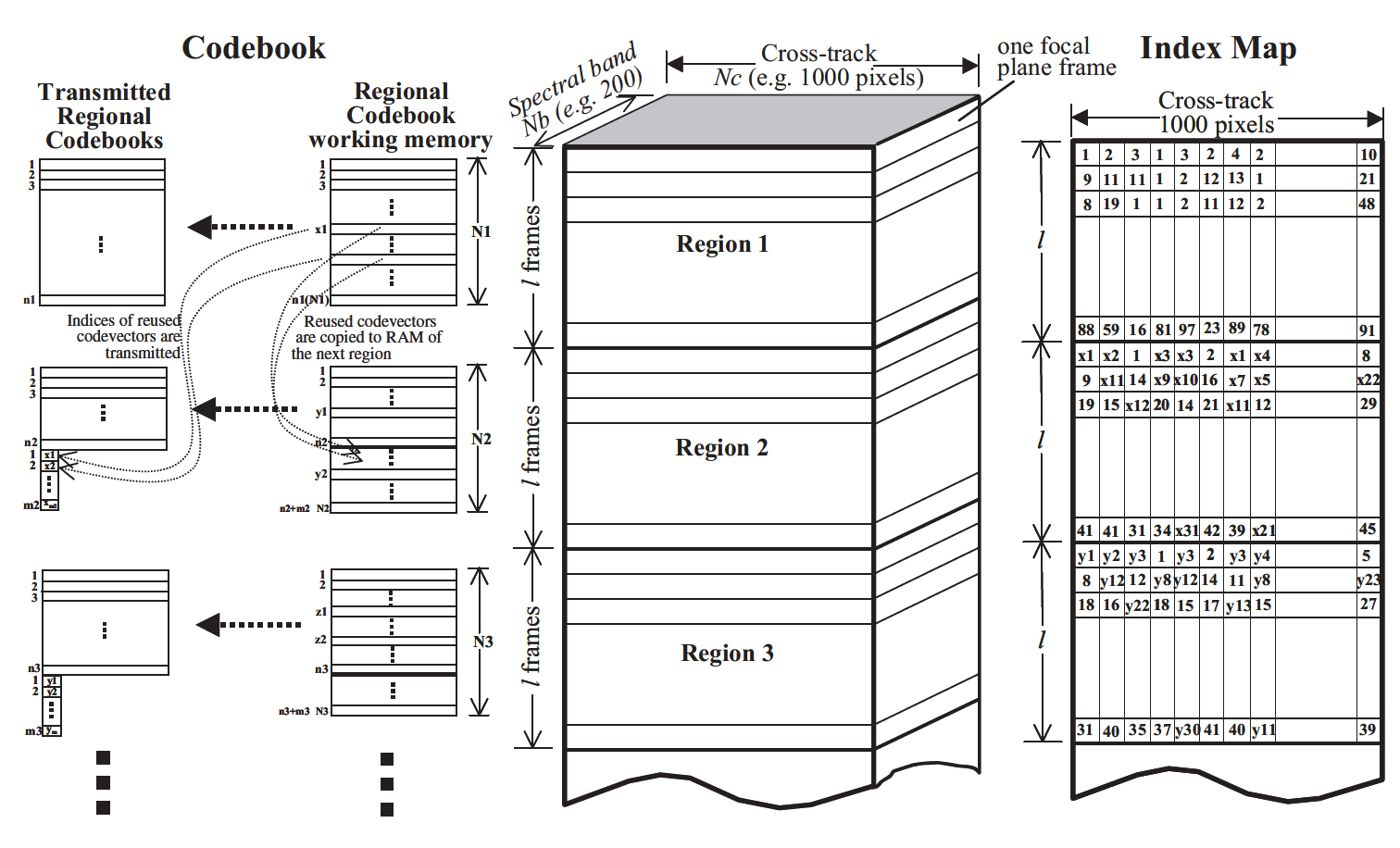}
   \end{tabular}
   \end{center}
   \caption[HSOCVQ] 
   { \label{fig:HSOCVQ} 
Schematic diagram of HSOCVQ method.\cite{qian_optical_2013}}
   \end{figure}

While SAMVQ and HSOCVQ have their benefits, they also have defects that need to be carefully considered.
One of the main defects of SAMVQ and HSOCVQ is the potential loss of important image details during the quantization process. This can result in a reduction of image quality and the loss of valuable information within the satellite imagery. Additionally, SAMVQ may struggle to effectively compress certain types of satellite images, leading to suboptimal compression ratios and potentially larger file sizes than desired.
Furthermore, SAMVQ may exhibit limitations in handling complex and dynamic scenes within satellite images, which can lead to artifacts and distortions in the compressed images. These defects highlight the need for ongoing research and development to address these limitations and to explore alternative compression techniques for satellite imagery.

\subsection{VAE for Satellite Image Compression}
\label{VAE for Satellite Imag compression}
There has been increasing interest in applying Variational Autoencoders to satellite image compression due to their ability to efficiently capture complex data distributions and generate realistic images. Recent studies have demonstrated the effectiveness of VAEs in compressing satellite images while preserving important spatial and spectral information. For example, Oliveira et al. \cite{alves_de_oliveira_reduced-complexity_2021}proposed a new architecture with reduced complexity based on the Ballé et al. \cite{balle_variational_2018} model, tailored for satellite image compression, which achieved significant improvements in compression performance compared to traditional methods. In particular, this model uses a scale hyperprior to enhance image reconstruction, which was reduced in complexity in the experiments to attain to limited hardware constraints on board satellites. Additionally, Guerrisi et al. \cite{guerrisi_convolutional_nodate} explored the use of a convolutional VAE with linear layer bottleneck for satellite image compression.
 Ma et al.\cite{ma_high-resolution_2023} proposed a new architecture for satellite video compression using a scale hyperprior, and Gaussian mixture quantization enabled by multiple VAEs encoders and decoders. 

Moreover, VAEs have shown promise in addressing challenges such as data heterogeneity and limited labeled training data in satellite image compression tasks. By leveraging the probabilistic nature of VAEs, researchers have been able to effectively model the complex and diverse nature of satellite imagery, leading to more accurate reconstructions and improved compression ratios. For example, La Grassa et al. \cite{grassa_hyperspectral_2022} have used VAEs for hyperspectral image compression. These advancements highlight the potential of VAEs for satellite image compression and demonstrate their adaptability to specific domain requirements. It is evident from the recent studies that Variational Autoencoders have emerged as a powerful tool for satellite image compression. These studies have demonstrated the effectiveness of VAEs in capturing complex data distributions and preserving important spatial and spectral information in satellite images. Although given the relevant infancy of the field, there is still room for improvement.

\subsection{VAE and Neural Compression}

Recent advances in machine learning have opened the door to new classes of compression algorithms, resulting in the creation of a category in itself. Neural compression is an umbrella term encompassing compression algorithms that rely on machine learning models in different capacities, heavily inspired by generative models such as GANs, VAEs, normalizing flows, and autoregressive models. Connections between these classes of machine learning models and standard data compression can be drawn both in lossless and lossy compression. For instance, like generative models, JPEG defines a probability distribution that represents assumptions about the latent space, and as in VAEs we can sample from this distribution to reconstruct the original images.

More generally the purpose of generative models, in the context of data compression, ties in with probabilistic modeling of data in the form of entropy coding. In simple term, this can be described as encoding more likely messages or images with fewer bits to decrease transmission cost. The theoretical limit of a codeword length of an optimal prefix-free code is described by Shannon's source coding theorem \cite{yang_introduction_2023}. The theorem states that when losslessly compressing data the entropy of the data is equal to the minimum number of bits required on average to encode data. In information theory, entropy is a measure of uncertainty about its outcomes, also called information content or surprise. Mathematically entropy is described as: 

\begin{equation}
\label{eq1}
H[X] = \mathbb{E}_{x\sim{}P}[-log_2P(x)]
\end{equation}
and formally the message length \textit{C(x)} of an optimal prefix-free code \textit{C} is bounded as:

\begin{equation}
\label{eq2}
H[X]\leq{}\mathbb{E}[|C(X)|]\leq{}H[X]+1
\end{equation}
  
   \begin{figure} [b]
   \begin{center}
   \begin{tabular}{c} 
   \includegraphics[height=5cm]{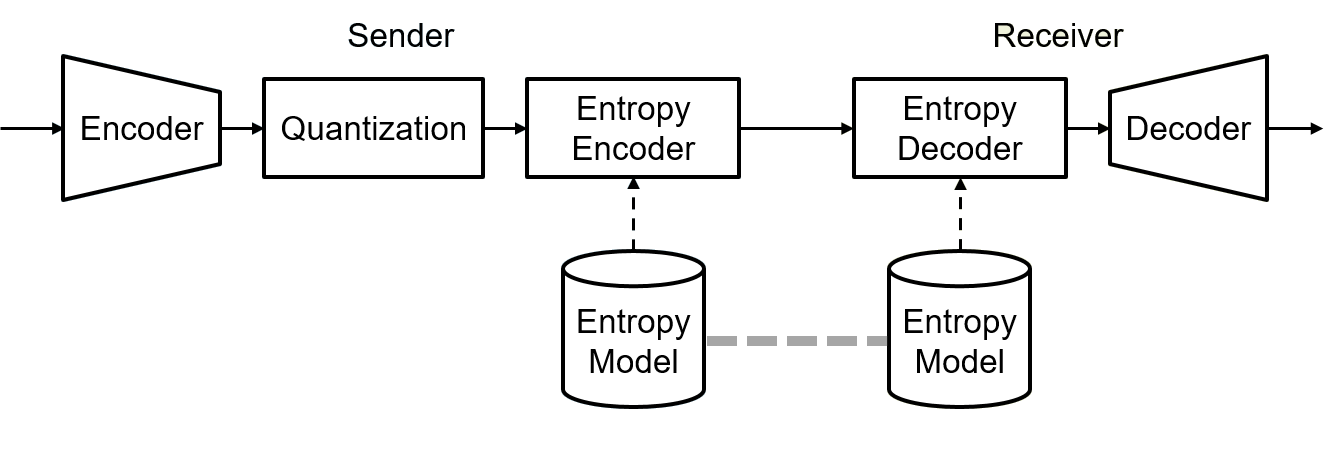}
   \end{tabular}
   \end{center}
   \caption[] 
   { \label{fig:HSOCVQ} 
Schematic diagram of image compression.}
   \end{figure} 

In practice we calculate the entropy of a distribution that represents the data we want to compress. Often we don't know the data distribution \textit{P}, or an intractable integral represents it. Therefore we approximate it using a distribution \textit{Q} by maximum likelihood or equivalent, minimizing the cross entropy (cross entropy defines how well Q(x) approximates P(X)), mathematically: 

\begin{equation}
\label{eq3}
H[\textit{P},\textit{Q}] = \mathbb{E}_{x\sim{}P}[-log_2Q(x)]
\end{equation}

From a data compression perspective, in the lossless case, we use a machine learning model to estimate data distributions to complement existing lossless methods such as arithmetic coding, asymmetric numeral systems (ANS), and bits back coding.

Neural compression can also be used for end-to-end lossy image compression. rate-distortion theory provides a fundamental framework in information theory that defines the theoretical limits of the performance of lossy compression algorithms. It explains the trade-off between the rate, which is the amount of compression, and the distortion, which is the measure of the loss in quality of the reconstructed signal compared to the original. Specifically, rate-distortion theory provides lower bounds on how much a signal can be compressed while maintaining a certain level of fidelity or quality.

The rate-distortion function specifies the minimum average rate for a given level of distortion, or conversely, the minimum distortion for a given rate; effectively generalizing Shannon's source coding theorem. In practice, however, achieving the exact rate-distortion bound is challenging, especially for complex data sources like images, and practical compression schemes typically operate above the theoretical bound. Mathematically the composite optimization objective is to simultaneously minimize the rate and the distortion defined as:

\begin{equation}
\label{eq4}
Rate:= \mathbb{E}[-log_2P([[f(X)]])]
\end{equation}

\begin{equation}
\label{eq5}
Distortion:= \mathbb{E}[[\rho(X,g([[f(X)]])]]
\end{equation}

Usually used in the form of a weighted composite objective or in the form of a rate-distribution Lagrangian (that uses a Lagrangian multiplier):

\begin{equation}
\label{eq6}
\mathbb{E}[-log_2P([[f(X)]])] + \lambda\mathbb{E}[[\rho(X),g([[f(X)]])]]
\end{equation}

   \begin{figure} [b]
   \begin{center}
   \begin{tabular}{c} 
   \includegraphics[height=10cm]{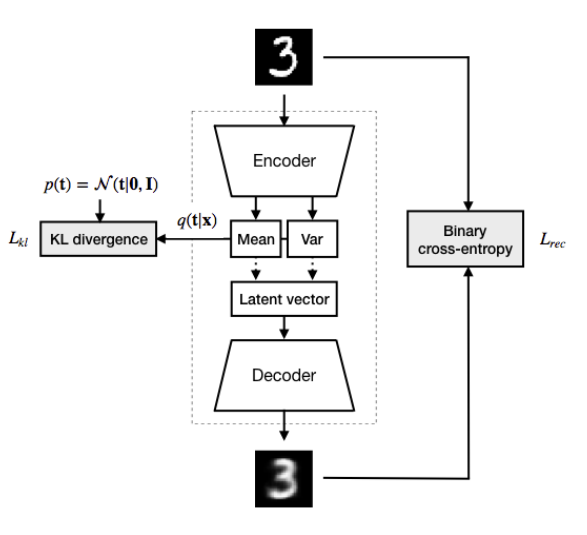}
   \end{tabular}
   \end{center}
   \caption[] 
   { \label{fig:VAE} 
Variational Autoencoder architecture schematic \cite{murphy_probabilistic_2022}.}
   \end{figure} 
Machine learning models, particularly those that learn end-to-end for a given distortion metric and rate constraint, attempt to approach this bound by using neural network architectures designed for compression. Rate-distortion optimization is often operationalized in the training of these models, where different points on the rate-distortion curve are obtained by adjusting hyperparameters that control the balance between the rate and distortion during the training process.

The rate-distortion objective resembles closely that of an autoencoder. Specifically variational autoencoders learn by approximating an expectation lower bound (ELBO) as a loss function. This ELBO is composed of a reconstruction term (MSE or BCE between original and reconstructed images) and a divergence term, describing the distance between the learned latent space distribution and the prior Gaussian distribution. The most popular VAE model is the $\beta$-VAEs which use Lagrangian multipliers to temper the divergence portion of the loss, similar to Equation \ref{eq6} is expressed as:

\begin{equation}
\label{eq7}
\mathcal{L} = \underbrace{\mathbb{E}_{q_{\phi}(z|x)}[log\hspace{0.1 cm} p_\theta(x|z)]}_\text{Distortion}+\underbrace{\lambda D_{KL}[q_\phi(z|x)||p(z)]}_\text{Divergence}
\end{equation}

When using VAEs for end-to-end lossy image compression we substitute the divergence loss component with a rate component such that the model learns the best compressed latent representations in terms of both compression rate (information content) and reconstruction quality. Then the unified objective becomes:

\begin{equation}
\label{eq8}
\mathcal{L} = \underbrace{\mathbb{E}_{q_{\phi}(z|x)}[log\hspace{0.1 cm} p_\theta(x|z)]}_\text{Distortion}+\underbrace{\lambda \mathbb{E}[-log_2P([[f(X)]])]}_\text{Bit-Rate}
\end{equation}

The latent space is then entropy-coded using the probability mass function shown in Equation \ref{eq10}, while during the training the non-differentiable rounding operation to the nearest integer is replaced with sampling from a box-shaped variational distribution defined as:

\begin{equation}
\label{eq9}
P(\hat{Z}=\hat{z}):= \int_{z_i-0.5}^{z_i+0.5}{\tilde{p}_{\theta{},i}(z_i) dz_i}
\end{equation}

\begin{equation}
\label{eq10}
z \sim{} Q_\phi{}(Z|X=x)
\end{equation}

\section{Methodology}

The methodology section of this study aims to provide a detailed description of the dataset and procedures employed to evaluate the effectiveness of Variational Autoencoders for satellite image compression. The section will outline the specific architectural modifications, encoding and decoding processes, as well as the evaluation metrics used to assess the performance of the VAE models in compressing satellite images.

\subsection{GRSS Dataset}

To study the image compression capabilities of the proposed model, the dataset provided for the IEEE 2023 GRSS data fusion contest and presented at CVPR 2022 by Huang et al.\cite{huang_urban_2022} was used. Images of the data fusion dataset were collected from the SuperView-1, Gaofen-2, and Gaofen-3 satellites, with spatial resolutions of 0.5 m, 0.8 m, and 1m, respectively. Data was collected from seventeen cities on six continents to provide a large and representative data set of high diversity regarding landforms, architecture, and building types.

The diverse data from the seventeen cities across six continents ensures that the dataset encompasses a wide range of landforms, architectural styles, and building types, making it highly representative of the complexities and heterogeneity present in satellite imagery. This diversity is essential for evaluating the adaptability of Variational Autoencoders to the challenging and varied characteristics of satellite images. The range of spatial resolutions and sources of the images, including those from SuperView-1, Gaofen-2, Gaofen-3, Gaofen-7, and WorldView, further contributes to the richness and complexity of the dataset, providing a comprehensive basis for assessing the performance of VAEs in compressing satellite imagery with varying levels of detail and spectral information.

   \begin{figure} [b]
   \begin{center}
   \begin{tabular}{c} 
   \includegraphics[height=6cm]{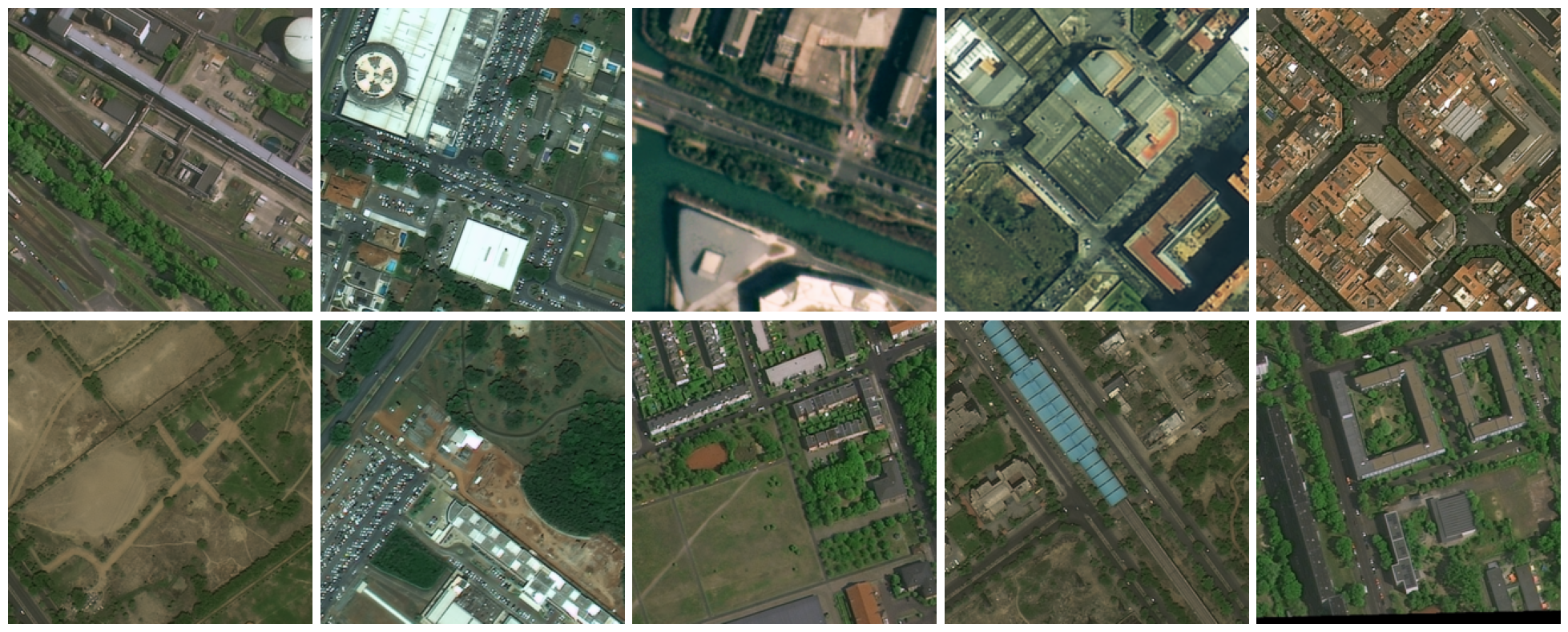}
   \end{tabular}
   \end{center}
   \caption[] 
   { \label{fig:GRSS} 
Samples from the GRSS dataset.}
   \end{figure}

\subsection{VAE Architecture}

The variational autoencoder architecture employed for the experiments is a classic VAE encoder-decoder structure, using integration as an approximation to the quantization losses in the entropy encoder layer. To maintain computational complexity as low as possible and flexible, the network is fully convolutional, does not have normalization layers, and does not employ the use of an external network to estimate hyperpriors as seen in other publications on satellite image compression discussed in Section \ref{VAE for Satellite Imag compression}. In this experiment, a classic VAE encoder-decoder structure is used for satellite image compression. The encoder part of the architecture takes satellite images as input and encodes them into a lower-dimensional latent space representation. The latent space is reparametrized to allow backpropagation and run through an entropy encoding layer, which also calculates the estimated bit rate of the latent concerning the original image. The decoder part of the architecture then takes the latent space representation and decodes it back into the original satellite image. This approach allows for the compression of satellite images while maintaining the essential features needed for downstream tasks such as image analysis and processing.

The experimental results of this approach demonstrate its effectiveness in compressing satellite images while preserving important details for subsequent analysis without using more complex architectures demonstrating that a 'vanilla' VAE still performs exceptionally well in the image compression task. By leveraging the capabilities of the classic VAE encoder-decoder structure, this method showcases a promising solution for satellite image compression with practical and computational advantages for hardware-constrained devices. 

   \begin{figure} [t]
   \begin{center}
   \begin{tabular}{c} 
   \includegraphics[height = 12.5 cm]{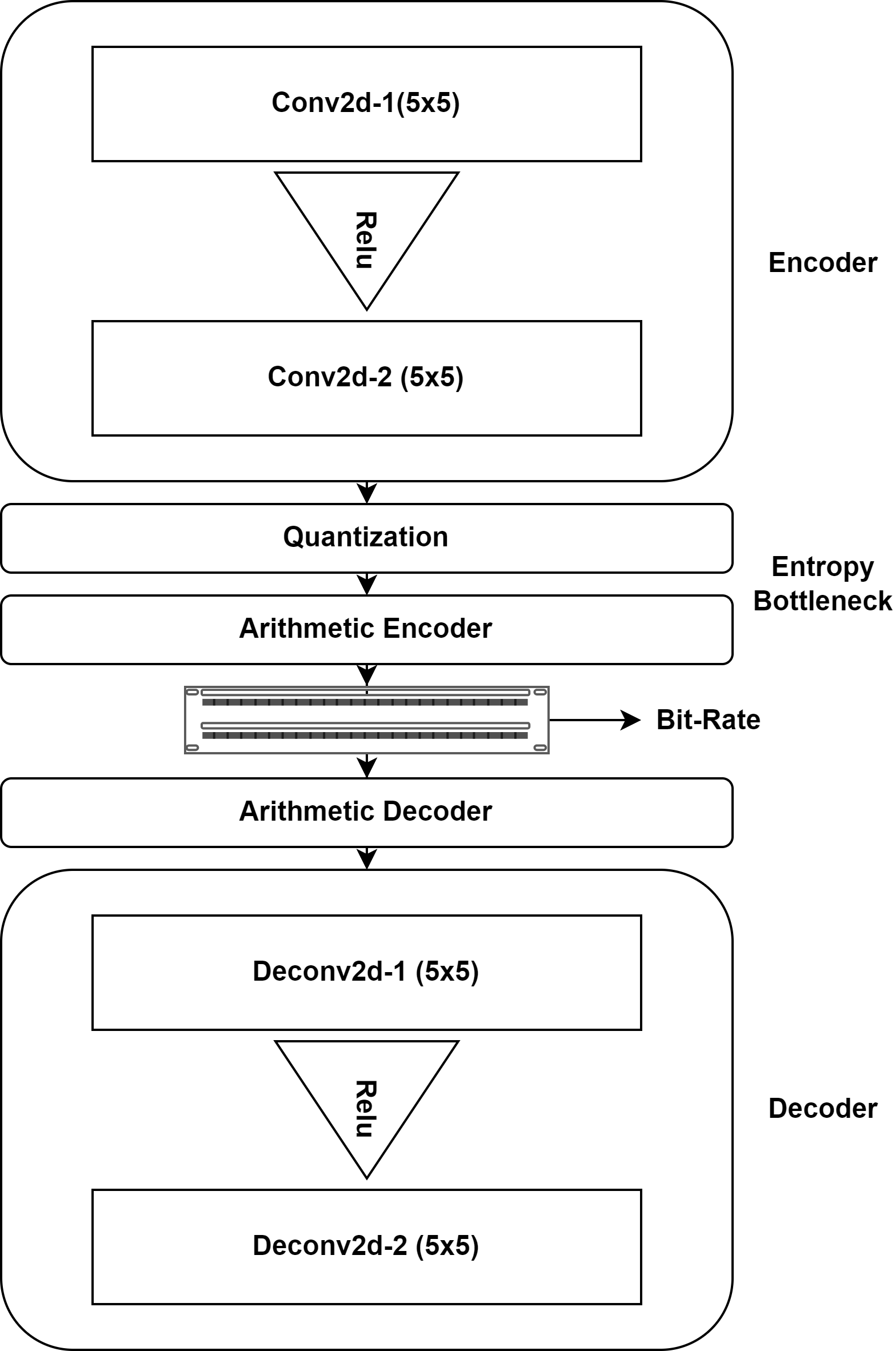}
   \end{tabular}
   \end{center}
   \caption[] 
   { \label{fig:architecture} 
VAE architecture used.}
   \end{figure}

\subsection{Evaluation Metrics}

The mean squared error (MSE) metric was used to evaluate the reconstruction distortion during training. To verify the quality of the reconstructed images two popular metrics were used on top of the MSE to estimate distortion, the structural similarity index (SSIM) and the peak signal-to-noise ratio (PSNR). SSIM is designed to improve on traditional metrics like MSE by considering changes in structural information rather than just pixel-level differences. It considers three comparison measurements between the samples of the two images: luminance which measures the closeness of the pixel intensities of the two images, contrast which compares the contrast of the two images by measuring the standard deviations of the pixel intensities, and structure which measures the correlation between the pixels of the two images. The SSIM combines these three factors into a single value that represents the similarity between the two images\cite{nilsson_understanding_2020}. The PSNR on the other hand quantifies the quality of the reconstruction by measuring the ratio of the maximum power of the original signal to the power of the noise that affects the quality of its representation. SSIM and PSNR are widely used in image processing and compression applications to assess the fidelity of the reconstructed images.
\begin{equation}
\label{eq11}
SSIM(x,y) = \frac{(2\mu{}_x\mu{}_y+c_1)(2\sigma{}_xy+c_2)}{(\mu{}_x^2+\mu{}_y^2+c_1)(2\sigma{}_x^2+\sigma{}_y^2+c_1)}
\end{equation}

\begin{equation}
\label{eq12}
PSNR(x,y) = 20log_{10}\left(\frac{MAX(x)}{\sqrt{MSE}}\right)
\end{equation}

\section{Results}

   \begin{figure} [b]
   \begin{center}
   \begin{tabular}{c} 
   \includegraphics[height = 5 cm]{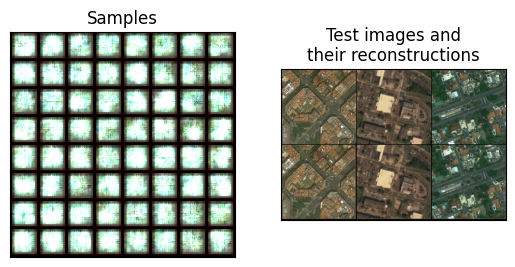}
   \end{tabular}
   \end{center}
   \caption[] 
   { \label{fig:reconstruction} 
Image reconstruction and sampling from latent space, sampling left, original image top right, reconstructed image bottom right, at latent space dimension= 128, hidden layer dimension = 256 (BPP = 4.71). The samples represent what the reduced dimensional representation of the original images looks like when sampling from the learned latent manifold.}
   \end{figure} 
   
   \begin{figure} [t]
   \begin{center}
   \begin{tabular}{c} 
   \includegraphics[width =16.7 cm]{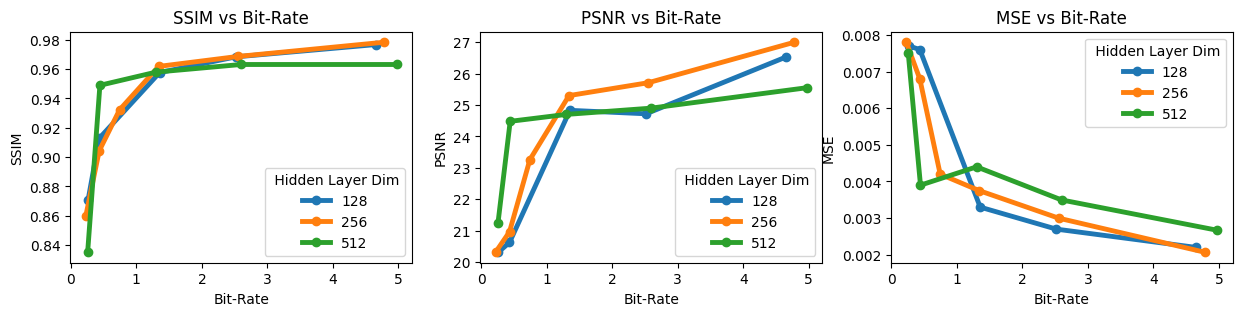}
   \end{tabular}
   \end{center}
   \caption[] 
   { \label{fig:plots} 
Image reconstruction metrics, SSIM left, PSNR middle, MSE right.}
   \end{figure}

The experimental results revealed that the VAE architecture achieved significant compression ratios while maintaining high-quality reconstructions, as evidenced by low MSE, high SSIM, and PSNR scores. The bit rate and entropy analysis further demonstrated the effectiveness of the compression approach in efficiently representing satellite images with minimal loss of information as shown in Figures  \ref{fig:reconstruction}, and \ref{fig:plots}.
The experiments were run at [4,8,16,32,64,128] dimensions of the latent space in the entropy bottleneck, which is how the bit-rate was controlled. Increasing the latent space dimensions also increased the bits being stored per pixel, resulting in better reconstruction quality at the cost of less compression, in line with the rate-distortion theorem. The theoretical limit of lossless compression based on Shannon's entropy theorem was calculated to be 0.9108, in comparison the lossy performance of the model achieves roughly 95\% structural similarity at that level, with very few model parameters and a relatively simple structure. Larger models achieve higher reconstruction quality although the increase is minimal compared to the increase in parameter size.
The beta lagrangian multiplier, as seen in Equation \ref{eq8} was selected to be 0.001 through trial and error testing, the focus being more on reconstruction quality. 

The results showcase that a very simple model with a few thousand parameters can visually reconstruct the image well enough to be of acceptable visual quality and achieve good compression rates. The good performance and relative simplicity show that these types of algorithms could be used by smaller satellites to effectively compress images. In addition, the scalability of this method makes it suitable for real-time processing on resource-constrained devices like unmanned aerial vehicles or edge computing systems. This presents exciting possibilities for deployment in practical scenarios where computational resources are limited while still requiring high-quality image reconstruction.

In comparison with hyperprior-enhanced methods, the reconstruction performance is slightly worse, as expected. Still, the tradeoff between computational expense and reconstruction quality highlights how the complexity scales exponentially when compared to compression performance. Therefore for applications of the nature of smaller satellites where the resources are heavily constrained simpler architecture can offer a better trade-off between quality and computational expense.

Furthermore, the compression scheme is inherently secure since the latent space prior is uniform and fully learned. In a satellite communication scenario the VAE would be trained a priori and then solely the encoder would be employed on board (further reducing and distributing computational expense). The decoder would be at the ground station trained to decode the messages. The system would be inherently secure because of the nonlinear transformations involved and the only way to decyper the communication would be to train a decoder with the same architecture to do that, which would need clear images to do so and implement a loss function for the optimization process and is therefore infeasible.

\section{Conclusion}

The experimental results demonstrate the success of the VAE architecture in achieving significant compression ratios while maintaining high-quality reconstructions. The low mean squared error scores, along with the high structural similarity index and peak signal-to-noise ratio values, indicate the effectiveness of the compression approach in efficiently representing satellite images with minimal loss of information. The scalability of this method makes it suitable for real-time processing on resource-constrained devices like unmanned aerial vehicles or edge computing systems. The security of the compression scheme is another notable aspect, owing to the uniform and fully learned latent space prior. The system's inherent security lies in the non-linear transformations and the need for a similar architecture and clear images to intercept the communication, making it a robust option for secure communication. Overall, the results indicate that smaller satellites could effectively use this type of algorithm for image compression, and its potential deployment in practical scenarios where computational resources are limited while still requiring high-quality image reconstruction is promising.

\bibliography{spieref} 
\bibliographystyle{spiebib} 

\end{document}